\documentclass[usenatbib]{mn2e}

\usepackage{times}
\usepackage{amssymb}
\usepackage{rotating}
\usepackage{xspace}
\usepackage{caption}
\usepackage{amsmath}
\usepackage{color}
\input{epsf}

\title[The broad band SEDs of four `hypervariable' AGN]
{The broad band SEDs of four `hypervariable' AGN}

\author[J.\,S. Collinson et al.]{James S. Collinson$^1$\thanks{Email: j.s.collinson@durham.ac.uk}, Martin J. Ward$^1$, Andy Lawrence$^2$, Alastair Bruce$^2$ \newauthor Chelsea L. MacLeod$^3$, Martin Elvis$^3$, Suvi Gezari$^4$, Philip J.\ Marshall$^5$ and Chris Done$^1$ \\
 \\
$^1$Centre for Extragalactic Astronomy, Department of Physics, Durham University, South Road, Durham, DH1 3LE, UK\\
$^2$Institute for Astronomy, SUPA (Scottish Universities Physics Alliance), University of Edinburgh, Royal Observatory, Blackford Hill, Edinburgh, EH9 3HJ, UK\\
$^3$Harvard-Smithsonian Center for Astrophysics, 60 Garden St., Cambridge, MA 02138, USA\\
$^4$Department of Astronomy, University of Maryland, Stadium Drive, College Park, MD, 20742-2421, USA\\
$^5$Kavli Institute for Particle Astrophysics and Cosmology, P.O. Box 20450, MS 29, Stanford, CA 94309, USA\\
}

\pagerange{\pageref{firstpage}--\pageref{lastpage}} \pubyear{2015}

\newcommand{\xmm}{{\it XMM}\xspace}
\newcommand{\xmmn}{{\it XMM-Newton}\xspace}

\newcommand{\galex}{{\it GALEX}\xspace}
\newcommand{\wise}{{\it WISE}\xspace}

\newcommand{\optxagnf}{{\sc optxagnf}\xspace}

\newcommand{\xspec}{{\sc xspec}\xspace}

\def\M_BH{$M_{\rm BH}$\xspace}
\def\a_OX{$\alpha_{\rm OX}$\xspace}

\def\rchi{{$\chi^{2}_{\rm red}$\xspace}}

\def\H0{{\rm ~km~s^{-1}~Mpc^{-1}}}

\def\eg{{e.g.\ }}

\def\ie{{i.e.\ }}

\def\la{\mathrel{\hbox{\rlap{\hbox{\lower4pt\hbox{$\sim$}}}{\raise2pt\hbox{$<$}}
}}}
\def\ga{\mathrel{\hbox{\rlap{\hbox{\lower4pt\hbox{$\sim$}}}{\raise2pt\hbox{$>$}}
}}}
\def\d25{$D_{25}$}
\def\nh{{$N_{\rm H}$}}
\def\Ha{{H$\alpha$}\xspace}
\def\Hb{{H$\beta$}\xspace}

\def\Heii{{He$\,${\sc ii}}\xspace}

\def\Mgii{{Mg$\,${\sc ii}}\xspace}
\def\Oiii{{[O$\,${\sc iii}]}\xspace}
\def\Feii{{Fe$\,${\sc ii}}\xspace}
\def\Sii{{[S$\,${\sc ii}]}\xspace}
\def\Nii{{[N$\,${\sc ii}]}\xspace}

\begin{document}

\maketitle

\label{firstpage}

\begin{abstract}
We present an optical to X-ray spectral analysis of four `hypervariable' AGN (HVAs) discovered by comparing Pan-STARRS data to that from SDSS over a 10 year baseline (Lawrence et al 2016). 
There is some evidence that these objects are X-ray loud for their corresponding UV luminosities, but given that we measured them in a historic high state, it is not clear whether to take the high-state or low-state as typical of the properties of these HVAs. We estimate black hole masses based on \Mgii and \Ha emission line profiles, and either the high and low state luminosities, finding mass ranges  $\log(M_{\rm BH}/M_{\odot}) = 8.2-8.8$ and $\log(M_{\rm BH}/M_{\odot}) = 7.9-8.3$ respectively. We then fit energy conserving models to the SEDs, obtaining strong constraints on the bolometric luminosity and \a_OX. We compare the SED properties with a larger, X-ray selected AGN sample for both of these scenarios, and observe distinct groupings in spectral shape versus luminosity parameter space. In general, the SED properties are closer to normal if we assume that the low-state is representative. This supports the idea that the large slow outbursts may be due to extrinsic effects (for example microlensing) as opposed to accretion rate changes, but a larger sample of HVAs is needed to be confident of this conclusion.
\end{abstract}


\section{Introduction} \label{sec:introduction}

\subsection{Background} \label{subsec:bg}

Based on variability, luminosity, and multi-frequency spectral energy distributions (SEDs), it is now accepted that gas accretion onto central galactic supermassive black holes (BHs) is the mechanism by which large amounts of energy are radiated from active galactic nuclei (AGN). We can probe these objects spectroscopically and temporally, as they emit across a great range of energies and exhibit variability at many frequencies (\eg \citealt{salpeter64}, \citealt{ward87}, \citealt{cristiani90}, \citealt{alexander12}, \citealt{collinson16}).

In an earlier paper (\citealt{lawrence16}, hereafter L16), we described the discovery of a large sample of ``hypervariable AGN (HVAs)''.
A search was made, originally for variable tidal disruption event (TDE; \eg \citealt{rees88}) candidates, in the Panoramic Survey Telescope And Rapid Response System (Pan-STARRS) database, which repeatedly surveys large regions of the sky with high cadence, and is therefore well-suited to searches for variable objects (\eg \citealt{gezari12}, \citealt{morganson15}). Pan-STARRS detections were compared to the $\sim$10 year earlier Sloan Digital Sky Survey (SDSS) measurements, and objects that had undergone a large increase in brightness ($|\Delta m| \! > \! 1.5$ mag in at least one optical filter) were selected as candidates. In addition to TDE and central region supernova (SN) candidates, this search yielded a significant number of very blue objects that follow-up photometry showed to be evolving on timescales of several years, whereas SNe and TDEs typically fade over time periods of weeks--months (\citealt{lawrence16}). Moreover, many of these slow, blue, variable objects were still increasing in brightness. Spectroscopy revealed them to be AGN at moderate redshifts ($z \! \sim \! 1$). Such extreme variability is rare for AGN, leading us to explore the possible explanations for these HVAs. Our definition of `hypervariable' here differs from that in \cite{morganson15}; that study defines all objects with $|\Delta m| \! > \! 2$ mag as `hypervariable', and as such includes a large number of highly variable stars and other phenomena, in addition to some AGN.

\subsection{Possible mechanisms for HVAs} \label{subsec:mechs_thispaper}

 L16 examined the properties of this growing sample of HVAs (currently 76 objects) and considered several interpretations of the data. These included highly luminous, slowly-evolving TDEs, line-of-sight extinction changes, extreme accretion rate changes and foreground microlensing.

L16 noted that the TDE explanation requires unusually massive stars ($\sim 10 \, M_{\odot}$) to be torn apart by the BH tidal forces to satisfactorily account for the observed event luminosities. Based on likelihoods and previously reported TDE candidates, disrupted stars are more likely to be of less than a solar mass ($\sim 0.3 \, M_{\odot}$), and occur around lower mass BHs ($\sim 10^6-10^7 \, M_{\odot}$) due to the steeper potential gradient (\citealt{gezari12}, \citealt{guillochon13}). This interpretation then seems quite unlikely in the whole sample of HVAs.

Extinction scenarios have been previously proposed to explain high amplitude variability in AGN, such as the transition from a Type 1 to Type 1.9 AGN for the object described in \cite{lamassa15}. L16 noted that whilst the timescale of such events makes them plausible candidates, there would be a strong colour change expected as the event evolves, which is not observed. A changing optical depth might produce such an effect, such as an eclipse by an opaque cloud. However, as many of the HVAs are observed to be decaying again, this model would consistently require successive extinction events.

The third possible origin of the variability considered in L16 is an accretion rate change. This is difficult to explore, as we do not yet have a model that adequately describes even conventional AGN variability (\eg \citealt{czerny04}, \citealt{lawrence12b}), but in this case, these HVAs could pose an intriguing means of probing extreme accretion rate changes.

The fourth possible origin considered by L16 is a foreground microlensing event. Such a scenario may arise from a star in a foreground galaxy passing between us and a background AGN, increasing the observed flux from the AGN by a large factor. L16 examined this hypothesis, finding that the timescales and rates expected are approximately consistent with those observed in HVAs. Some of the long-term light curves look like the symmetric, cuspy shapes na\"ively expected from point source-point lens models (\eg \citealt{schmidt10}), or the double peaked structure expected from binary lenses or lenses sheared by the parent galaxy of the lensing star (\eg \citealt{chang84}). Others show erratic light curves which probably are not caused by microlensing. \cite{bruce16}, after a detailed treatment of a simple microlensing scenario, show that this is a plausible explanation for at least a subset of these objects.

Overall, one could make a distinction between models where the large outbursts are intrinsic, such as accretion rate changes, and those where the cause is extrinsic, such as extinction or microlensing.

\subsection{Aims of this paper}

In this paper we report the optical spectra and light curves, X-ray spectra and SEDs of four HVAs for which we have obtained \xmmn observations. We obtain BH mass estimates for each source from the \Mgii emission line, and fit SED models to constrain the mass accretion rates. Our primary aim is simply to examine the broad-band properties of the objects in order to establish whether the HVAs appear unusual when compared to other AGN. We do not aim to explicitly test either the accretion instability or microlensing models, or other specific models. Rather, we examine the observations in the context of the likely differences between possible intrinsic and extrinsic explanations in general.

\begin{table*}
\caption{\small The names (used in this paper), positions, redshifts, \xmm observation IDs and observation dates for the sample of four HVAs.}
\small
\centering
\begin{tabular}{cccccc}
\hline
Name & R.A. (J2000) & Dec. (J2000) & $z_{\rm meas}^a$     & \xmm Obs ID(s)       &
Obs UT(s) \\
\hline
J0312$+$1836  & 03 12 40.86 & $+$18 36 41.1  &  0.889     & (0724440) 101 \& 601 &  2013-08-12 \& 2014-01-22 \\
J1422$+$0140  & 14 22 32.45 & $+$01 40 26.7  &  1.078     & (0724440) 301$^b$ \& 801 &  2014-01-04 \& 2014-02-06 \\
J1519$+$0011  & 15 19 43.99 & $+$00 11 47.4  &  0.530     & (0724440) 401 \& 901 &  2014-01-27 \& 2014-02-10 \\
J2232$-$0806  & 22 32 10.51 & $-$08 06 21.2  &  0.276     & (0724441) 001 \& 101$^b$ &  2013-12-14 \\  

\hline
\multicolumn{6}{l}{$^a$Redshift measurement described in Section \ref{sec:bhmasses}.}\\
\multicolumn{6}{l}{$^b$Obs IDs 0724440301 and 0724441101 did not yield useful data, possibly due to background activity.}\\

\end{tabular}
\label{tab:sample}
\end{table*}

\begin{figure*}
\centering
\begin{tabular}{cc}
 \includegraphics[width=0.45\textwidth, angle=0]{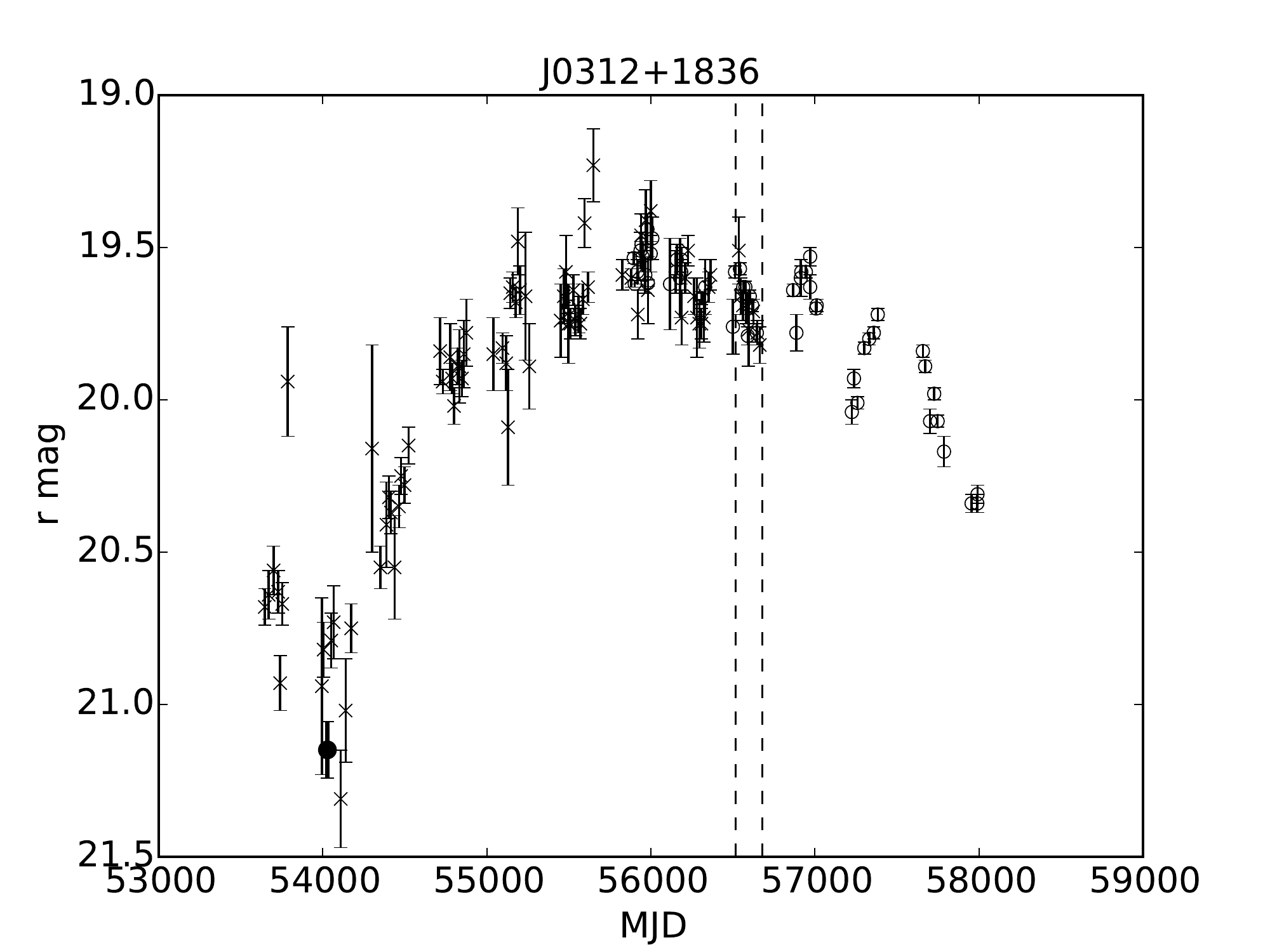}  &
 \includegraphics[width=0.45\textwidth, angle=0]{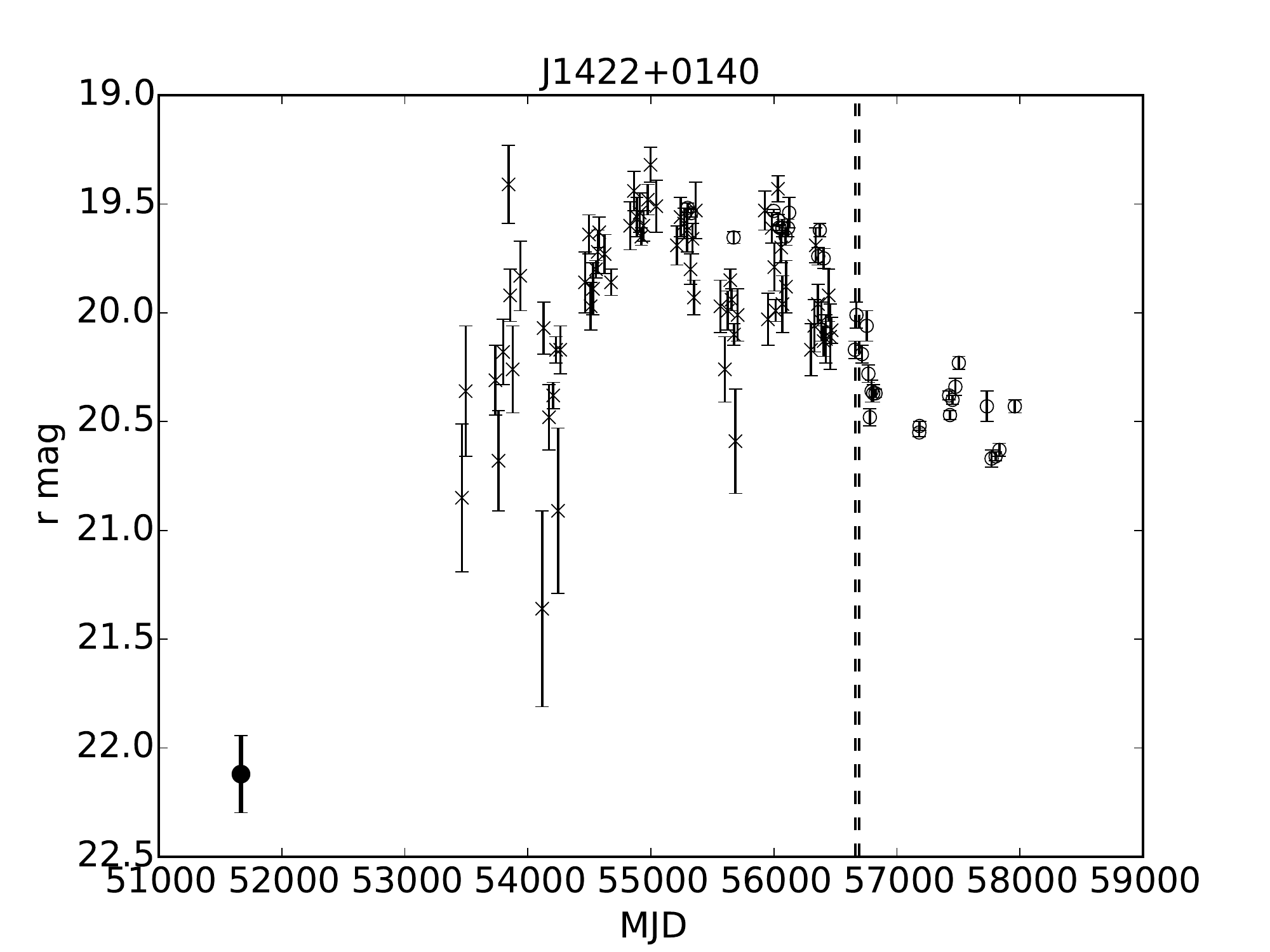} \\ 
 \includegraphics[width=0.45\textwidth, angle=0]{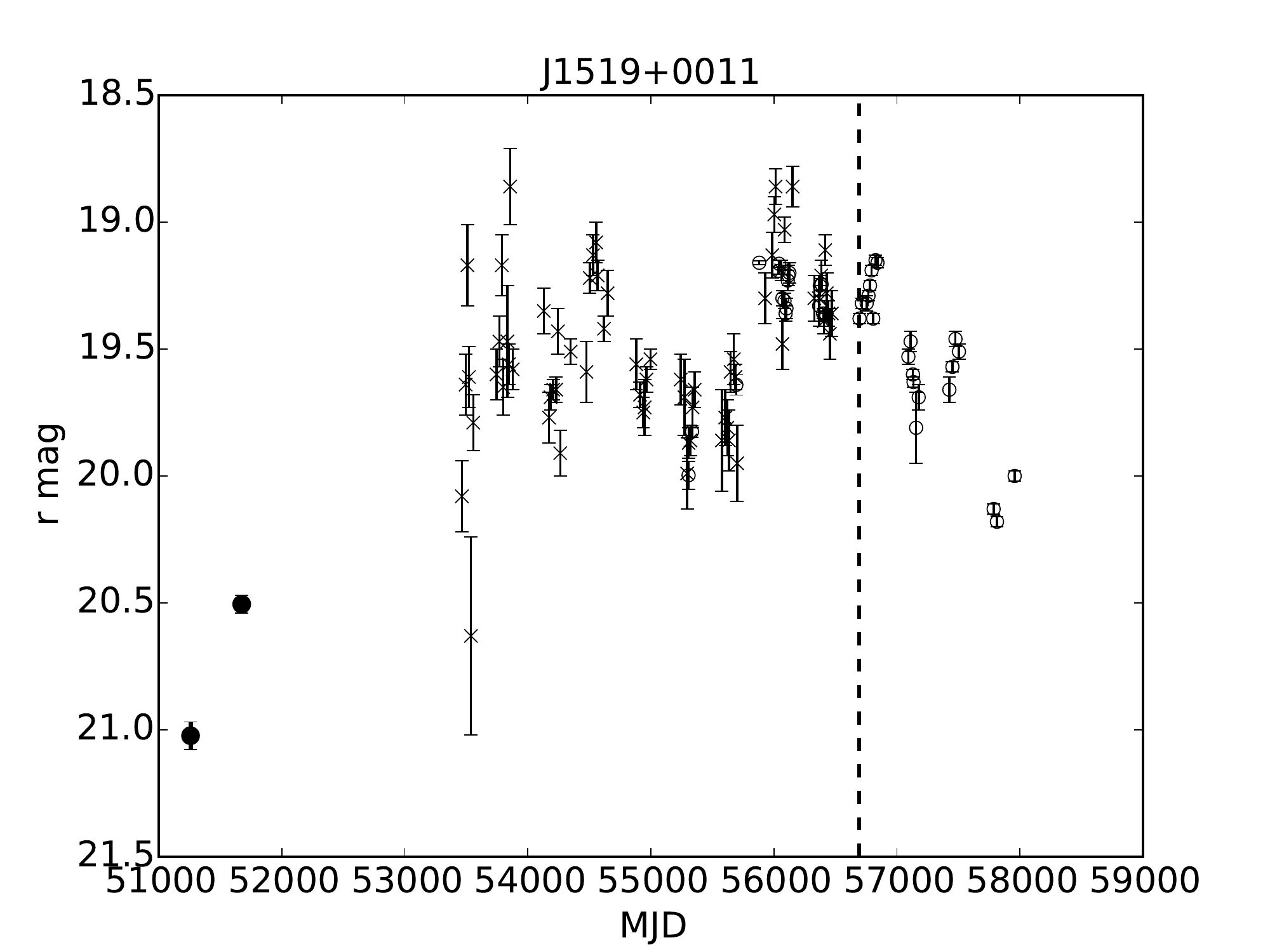}  &
 \includegraphics[width=0.45\textwidth, angle=0]{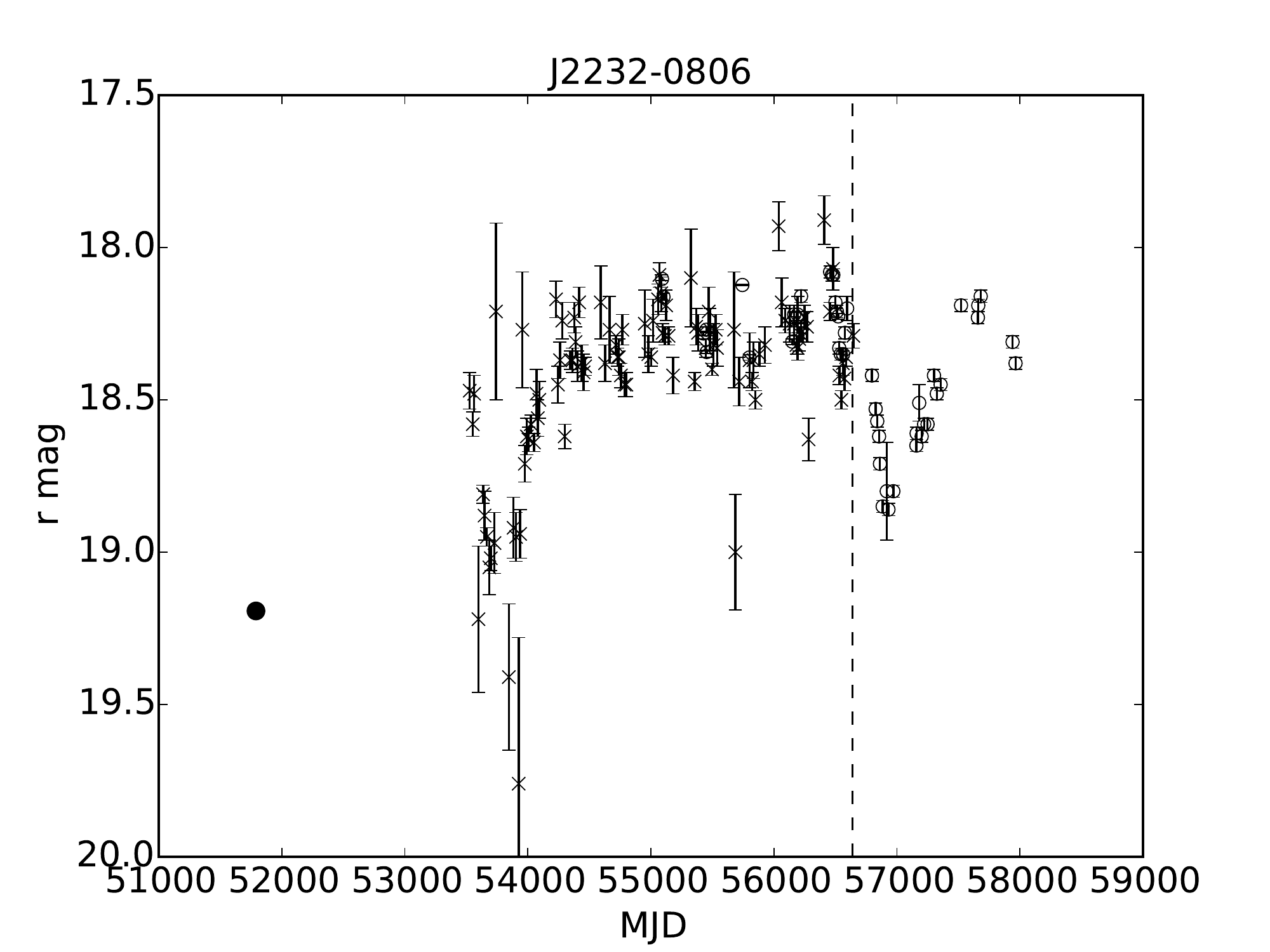} \\ 
\end{tabular}
\caption{\small The long-term light curves for the four HVAs. The open circles are PanSTARRS and LT data; the solid circles are SDSS data; the crosses are CRTS data. The dates of the \xmm observations are also marked, and shows that there is likely to be only a small, if any, change in flux between observations, for those objects that were observed on two different epochs.}
\label{fig:lightcurves}
\end{figure*}

\section{Sample and observations}  \label{sec:sample}

Pan-STARRS 1 (PS1) is a 1.8 m wide-field telescope, situated on the Hawaiian island of Maui, that was originally designed for the detection of near-Earth objects \citep{kaiser04}. It saw first light in 2007, and the science mission commenced in 2010. Since March 2010, it has surveyed the sky in five photometric bands ($g_{\rm P1}$, $r_{\rm P1}$, $i_{\rm P1}$, $z_{\rm P1}$ and $y_{\rm P1}$); 56 per cent of the observing time is devoted to the 3$\pi$ survey, with additional, deeper observations of smaller sky regions making up the `Medium Deep Survey' (\citealt{kaiser10}, \citealt{magnier13}).

The means of  selecting our full sample of transients is described in detail in L16. In summary, the parent sample was selected by comparing PS1 magnitudes in the Faint Galaxy Supernova Search (FGSS) database \citep{inserra13} to earlier SDSS DR7 observations (\citealt{york00}, \citealt{abazajian09}). SDSS detections were required to be coincident in position to within 0.5 arcsec, classified as galaxies, and show a $\geqslant$1.5 mag increase in brightness when reobserved by PS1.

The SDSS and PS1 FGSS photometry were supplemented with existing data from the Catalina Real-time Transient Survey (CRTS -- \citealt{drake09}) when available. Pointed follow-up observations with the Liverpool Telescope (LT) at the Roque de los Muchachos Observatory on La Palma were also made, starting in 2011. This additional photometry was used to reject probable supernovae by applying both colour cut and decay time criteria (see also \citealt{lawrence12a}). This results in an ever-growing sample of HVAs, with 76 discovered so far.

Optical spectra were obtained using the William Herschel Telescope Intermediate dispersion Spectrograph and Imaging System (WHT/ISIS) for much of the sample over several epochs. The data collection and reduction is described in L16 and \cite{bruce16}. Spectra were normalised to the LT $g$ band light curve, to mitigate uncertainty in the flux calibration.

We observed four representative, but bright, HVAs with \xmmn during August 2013 to February 2014, and this forms the sample examined in this paper. The objects' names, positions, redshifts and \xmm observation details are listed in Table \ref{tab:sample}. Where possible, observations were split between two epochs to look for variability.

The \xmmn pipeline processing system (PPS) products were extracted from the \xmm science archive (XSA). The PPS uses tasks from the science analysis system to extract events, using only `good' observation time where the background effects and flares were minimal. Our objects are not extended, and are not bright enough for pile-up to occur. A fuller discussion of the PPS data reduction procedure is given in \cite{watson09}.

Throughout this paper we will use all available data for each object from the European Photon Imaging Camera (EPIC) MOS1, MOS2 and PN detectors. This will maximise the SNR of the X-ray spectrum.

\section{Light curves}  \label{sec:lightcurves}

Long term light curves are shown in Fig. 1. Two of the objects (J0312 and J1422) show slow smooth outbursts by 1-3 mags, with superimposed variability. 
J1519, and possibly J1422, show evidence for two (or more) peaks. This is expected in a binary, or sheared ``Chang-Refsdal'', microlensing scenario, though it is not our intention to test this hypothesis at this stage. 
The key point is that for all of J0312, J1422, and J1519 it looks like the low-state may be the normal one. The fourth object (J2232) looks rather different, showing a flat top with dips. In this case, it may be that the high state is the normal one. 

\begin{table*}
\caption{\small The key $r$ band magnitudes for the four HVAs, including the faint state, and at the times of \xmm observation. The means by which we measure each, as discussed in the text, is shown in brackets. We also show $\Delta m$, the magnitude difference between \xmm observation and quiescent state. This is the difference in brightness between the two scenarios we test in this paper.}
\small
\centering
\begin{tabular}{ccccccc}
\hline
Name 
& $m_{\rm min,SDSS}$  
& $m_{(X\!M\!M \, 1)}$  
& $m_{(X\!M\!M \, 2)}$ 
& $\Delta m =$\\

& (SDSS)  
& (Interpolation) 
& (Interpolation) 
& $m_{\rm min,SDSS}-m_{(X\!M\!M \, 1)}$\\
\hline
J0312$+$1836  
& $21.26\pm0.06$ 
& $19.62^{+0.07}_{-0.04}$ 
& $19.72^{+0.04}_{-0.05}$ 
& 1.64 \\
J1422$+$0140  
& $22.12\pm0.18$ 
& $20.10^{+0.07}_{-0.10}$ 
& $20.12^{+0.06}_{-0.10}$ 
& 2.02 \\
J1519$+$0011  
& $21.02\pm0.05$ 
& $19.35^{+0.03}_{-0.04}$ 
& $19.35\pm0.03$ 
& 1.67\\
J2232$-$0806  
&$19.193\pm0.016$
& $18.31\pm0.06$ 
& -- 
& 0.88 \\
\hline
\end{tabular}
\label{tab:mags}
\end{table*}

For this study, we use spectral data from a range of dates to estimate the BH mass (\M_BH), and then fit the broad-band SED with an energy conserving accretion model \citep{done12}. However, as these HVAs are variable, we first require estimates of the largest and smallest magnitudes, corresponding to the faint and bright states respectively. For this we use the $r$ band, as it is close to the middle of the spectral range, and corresponds to the same, or a very similar bandpass in PS1, LT and SDSS. We shift the CRTS magnitudes to match the SDSS/PS1/LT $r$ magnitudes, as observations made by CRTS are in white light, calibrated to a $V$ band zero point.

We wish to make an estimate of the magnitude of each HVA at the time of the \xmm observation. As LT photometry exists both
prior and subsequent to the XMM observations, we
linearly interpolate these observations to estimate the magnitudes at
these epochs. We estimate the
error using a bootstrap technique, whereby 1000 random datasets
are drawn from the LT data. These are presented in Fig. 2. We tabulate
the $r$ band magnitudes from this analysis in Table \ref{tab:mags}.

\begin{figure}
\centering
\includegraphics[width = 0.5 \textwidth]{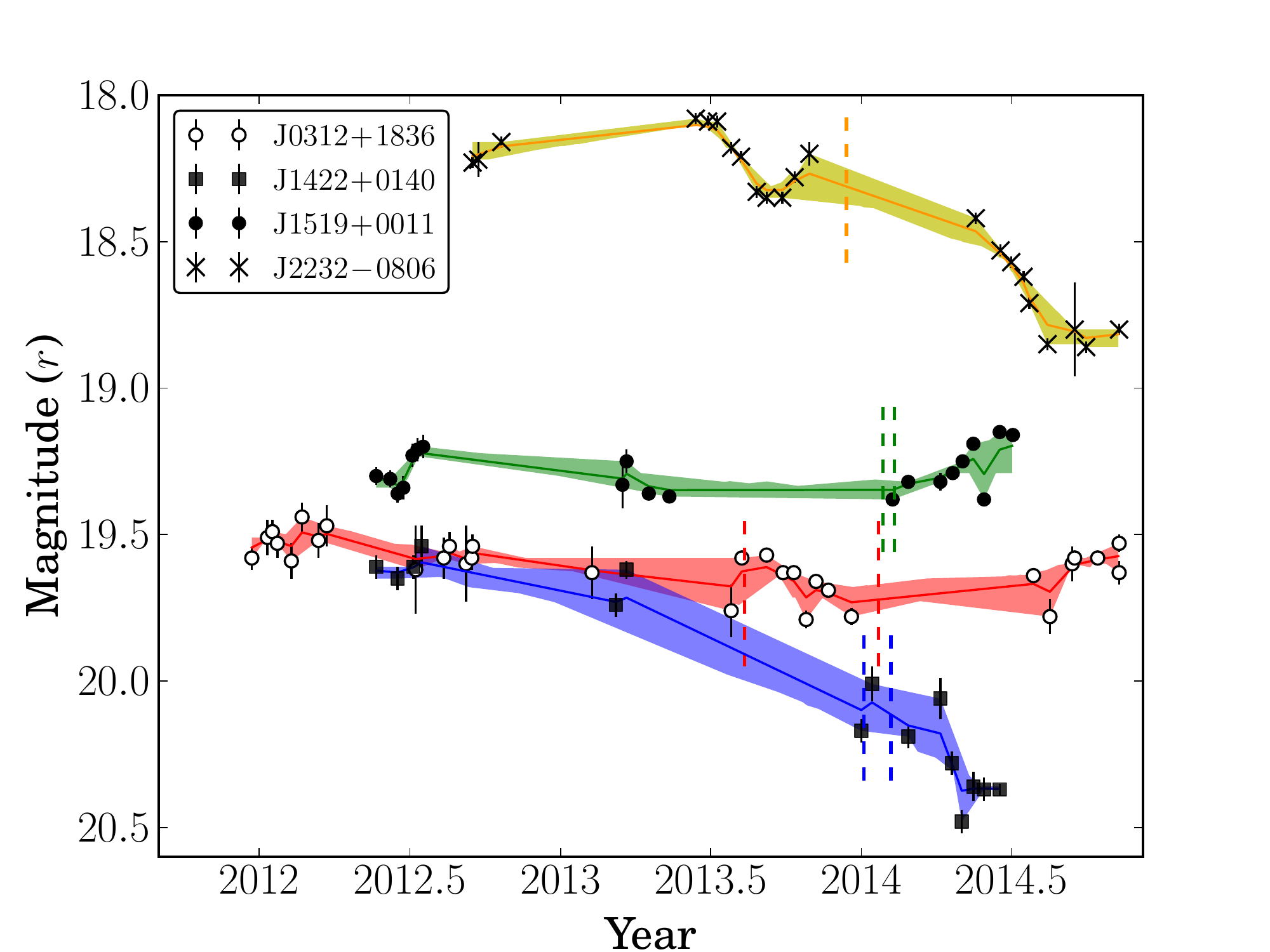}
\caption{\small The LT light curves for the four HVAs. We use linear interpolation to measure the magnitude at the epoch of \xmm observation, and utilise a bootstrap subsampling technique to estimate the 1$\sigma$ error (shaded regions) on these measurements. The dates of the \xmm observations are again marked by vertical dashed lines. On this scale, it appears there is statistically significant intrinsic variability, on top of the global variability, in some objects.}
\label{fig:lightcurves_xmm}
\end{figure}

In Fig.\ \ref{fig:lightcurves_xmm} we do see evidence for statistically significant variability on top of the global variability trend in some objects. However, for objects with two \xmm observations, there is no evidence, from the optical lightcurve, for a change in magnitude between the two \xmm observation dates (Table \ref{tab:mags}).

\section{Black hole masses} \label{sec:bhmasses}

\begin{figure}
	\centering
	\includegraphics[trim={0 1.3cm 0 1.51cm},clip, width = 0.5 \textwidth]{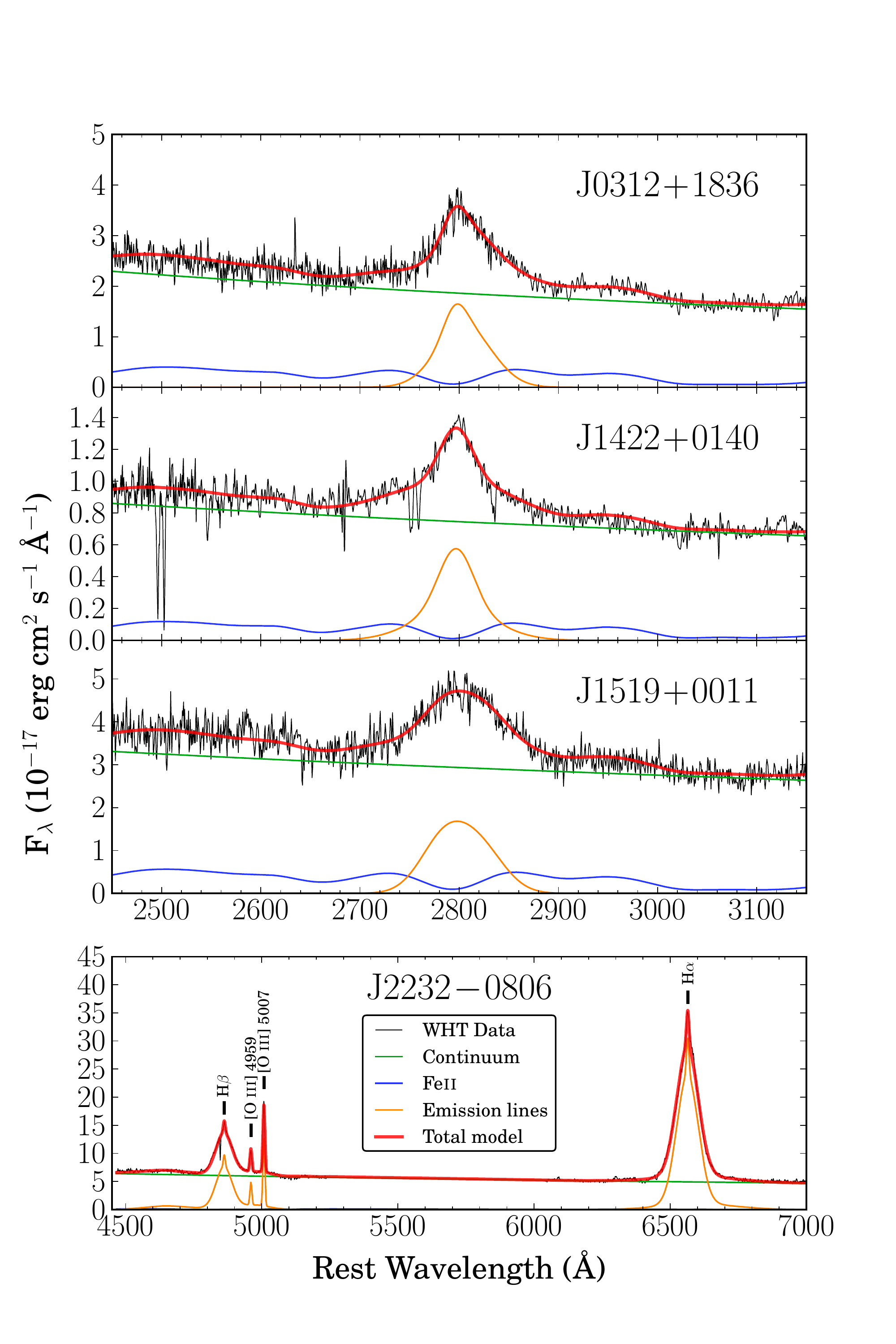}
	\caption{\small Examples of our model fits to the optical spectrum of each object. In black is the data from WHT/ISIS, and red is the fitted model profile, with constituent components shown. The top three panels show the higher redshift objects, for which only \Mgii is available for fitting, and the bottom panel shows J2232$-$0806, in which we can model \Ha and \Hb , in addition to the narrow \Oiii doublet; these are marked. We fitted these models to every spectrum for each object (across multiple dates) and took the average as our best estimate. In the bottom panel, we masked out a region of the spectrum between \Ha and \Hb , as it contained telluric sky features that would have affected the fitting. The cases shown here are for the spectra normalised to the quiescent state.}
	\label{fig:masses}
\end{figure}

The optical SED of AGN matches well the emission expected from a geometrically thin, optically thick accretion disc (AD) of gas (\eg \citealt{shakura73}, \citealt{davis07}). The energy output is thermally dominated; each radius emits blackbody radiation of characteristic temperature that increases to smaller radii, and the peak disc temperature depends on \M_BH, mass accretion rate and spin. Modelling the SED therefore requires an estimate of \M_BH.

This can be accomplished by several methods, the most accurate of which is reverberation mapping (RM) (see \citealt{blandford82} for original description, and \eg \citealt{peterson04}, \citealt{denney10} and \citealt{du14} for applications). However, RM requires long-term spectroscopic monitoring and thus we are unable to apply it to our sample. Instead, we base our \M_BH estimates on single-epoch observations of the broad emission lines, using methods calibrated from RM. This approach requires the linewidth to obtain a Keplerian velocity, and a simultaneous luminosity, which determines the radius at which the line is emitted (the so-called broad line region (BLR) size). The orbital velocity at that radius then implies a central mass (\eg \citealt{wandel99}, \citealt{kaspi00}, \citealt{vestergaard02}, \citealt{bentz06}).

As we only have WHT/ISIS spectra covering optical wavelengths, we must use the \Mgii line in the three highest redshift objects, and apply the method of \cite{mclure04}. For the lower redshift J2232$-$0806, we can make an estimate of \M_BH using the well-studied Balmer emission lines, \Ha and \Hb , and the methods of \cite{greene05} and \cite{woo02} respectively.

We correct these spectra for Milky Way (MW) reddening, using the dust maps of \cite{schlegel98} and the extinction law of \cite{cardelli89}, and then decompose the broad emission lines using the following standard procedures, as described in \eg \cite{greene04}, \cite{wang09}, \cite{shen12} and \cite{matsuoka13}. The underlying optical continuum in AGN closely approximates to a power-law over limited wavelength ranges, and the profiles of the broad emission lines can be modelled by a combination of Gaussian components. The blended \Feii emission observed in many AGN can also be modelled using templates derived from local AGN.

As these HVAs have by definition undergone a significant increase in brightness not typical of AGN, determining the luminosity for the \M_BH calculation is model dependent. On the one hand, we could assume that the observed (high) flux is the intrinsic level, and the corresponding continuum/line luminosity gives the best representation of the BLR size. For this case, we therefore use the WHT spectra as observed, applying no scaling factor. Alternatively, if the variability is caused by some extrinsic factor then the normal continuum level would be below that observed. For this situation, we should calculate the continuum/line luminosity from the spectra normalised to the HVA quiescent state, taken to be represented by the SDSS magnitude. Table \ref{tab:mags} illustrates the agreement between these two methods.

\begin{table*}
\caption{\small The key line properties for the sample, including FWHM, continuum/line luminosity and inferred \M_BH (according to equations \ref{eq:mgiimass}, \ref{eq:hamass} and \ref{eq:hbmass}). The error on the \M_BH is a measurement error only, and does not reflect systematic errors and uncertainties inherent in the method. We present \M_BH values resulting from analysing the spectrum as observed (used in scenario A) and from normalising the spectra to the quiescent state (SDSS) magnitude (used for scenario B).}
\small
\centering
\begin{tabular}{ccccccccc}

\hline
Name & Line & FWHM                      & $\lambda L_{\lambda}$ or $L_{\rm line}$ &
$\lambda L_{\lambda}$ or $L_{\rm line}$ & $M_{\rm BH,observed}$    & $M_{\rm BH,faint}$ \\
     &      & [km s$^{-1}$]             & [$\log $(erg s$^{-1}$)]                 &
[$\log $(erg s$^{-1}$)]                 & [$\log (M_{\odot})$]  & [$\log (M_{\odot})$] \\
 & & & (observed; A) & (quiescent; B) & (A) & (B) \\
\hline
J0312$+$1836  & \Mgii & $6000\pm400$ & $45.09\pm0.07$ & $44.330\pm0.011$ & $8.73\pm0.03$ & $8.26\pm0.06$  \\
J1422$+$0140  & \Mgii & $4800\pm400$ & $45.01\pm0.17$ & $44.146\pm0.005$ & $8.47\pm0.10$ & $7.95\pm0.08$ \\
J1519$+$0011  & \Mgii & $8100\pm200$ & $44.79\pm0.07$ & $44.004\pm0.013$ & $8.81\pm0.07$ & $8.33\pm0.03$ \\
J2232$-$0806  & \Ha   & $4350\pm70$  & $43.08\pm0.12$ & $42.76\pm0.08$   & $8.20\pm0.05$ & $8.03\pm0.04$ \\
              & \Hb   & $4350\pm70$  & $44.20\pm0.15$ & $43.852\pm0.011$ & $8.08\pm0.10$ & $7.86\pm0.02$ \\
\hline

\end{tabular}
\label{tab:masses}
\end{table*}

We then fit the following components:

\begin{itemize}
\item[i.]{The continuum underneath the emission lines is modelled as a power law of the form $F(\lambda) = C_1(\lambda / 5100 \, \textrm{\AA})^{-C_2}$, where $C_1$ and $C_2$ are free constants representing the normalisation and slope, respectively. We do not model the Balmer continuum, which could contribute to the continuum under \Mgii, as we only use a small part of the spectrum around the \Mgii line itself, and the power-law approximation is sufficient.}
\item[ii.]{The blended \Feii emission is modelled using the template of \cite{veroncetty04} in the optical, and \cite{vestergaard01} in the UV, both of which are derived from studies of the Type 1 AGN I Zwicky 1. This component has two free parameters: the normalisation and width of the convolving Gaussian.}
\item[iii.]{Emission lines are modelled as a sum of Gaussians. In the first three objects, of higher redshift, we model only \Mgii , with two Gaussian components. We do not attach a physical significance to these components and do not try to model the \Mgii line as a doublet, for the same reason as \cite{shen12}; the line splitting is too small to be significant. For J2232$-$0806, we model \Ha and \Hb with three components each (one narrow and two broad), which are locked together in velocity width and amplitude ratio. We also model the narrow \Oiii doublet with two components for each member, and the two lines are fixed at a 2.98:1 ratio \citep{storey00}. Finally we model \Heii with one component. We do not model the narrow \Nii doublet that is often seen on top of the \Ha profile, as there is no detected \Sii doublet redwards of \Ha, which indicates that \Nii contribution will be similarly negligible. The narrow components in \Ha and \Hb are fixed to the same velocity width as \Oiii .}
\end{itemize}

Redshifts were measured from the \Mgii (J0312$+$1836, J1422$+$0140 and J1519$+$0011) and strong \Oiii lines (J2232$-$0806). These are given in Table \ref{tab:sample}. \M_BH is then calculated according to the following equations. We use the relation in \cite{mclure04} for \Mgii (in the first three, higher redshift, objects):
\begin{equation} \label{eq:mgiimass}
M_{\rm BH} = 3.2 \times \left( \frac{\lambda L_{\rm \lambda}}{10^{44} \, \rm erg \, s^{-1}} \right)^{0.62}
\left( \frac{\rm FWHM_{MgII}}{\rm km \, s^{-1}} \right)^{2} M_{\odot}
\end{equation}
with $\lambda L_\lambda$ being the monochromatic continuum luminosity at 3000~\AA .

In J2232$-$0806, for \Ha , we use the method described in \cite{greene05}:
\begin{equation} \label{eq:hamass}
\begin{split}
M_{\rm BH} & = (2.0^{+0.4}_{-0.3}) \times 10^6 \\
 & \times \left( \frac{L_{\rm H \alpha}}{10^{42} \, \rm erg \, s^{-1}} \right)^{0.55 \pm 0.02} \! \!
\left( \frac{\rm FWHM_{H \alpha}}{10^3 \, \rm km \, s^{-1}} \right)^{2.06 \pm 0.06} \! \! M_{\odot} .
\end{split}
\end{equation}

For comparison purposes, in J2232$-$0806, we also use the \cite{woo02} method of measuring mass from \Hb , as an additional check of systematic uncertainties in the value:
\begin{equation} \label{eq:hbmass}
M_{\rm BH} = 4.817 \times \left( \frac{\lambda L_{\rm \lambda}}{10^{44} \, \rm erg \, s^{-1}} \right)^{0.7} \left( \frac{\rm FWHM_{H \beta}}{\rm km \, s^{-1}} \right)^2 M_{\odot}
\end{equation}
where $\lambda L_\lambda$ here is the monochromatic continuum luminosity at 5100~\AA .

Multiple spectra were available for each object (between three and five), and we performed our spectral analysis on each, measuring \M_BH from each spectrum independently and taking the mean as our best estimate. The standard deviation of these provides an estimate of the measurement uncertainty, which, it should be noted, is not the dominant source of error on such estimates. Example spectral decompositions are shown in Fig.\ \ref{fig:masses}, and the resulting \M_BH estimates are tabulated, together with observation dates, in Table \ref{tab:masses}. We use a Levenberg-Marquardt minimisation algorithm to fit the data, and employ a sigma-clipping routine in J1422$+$0140 to reduce the effect of the narrow absorption features we observe in that spectrum. As discussed above, we tabulate \M_BH values for two cases: one in which the spectra were taken as observed, and the other in which the spectra were scaled to the faint state. 

In principle, the two different approaches for estimating \M_BH could provide additional diagnostics. \M_BH is constant between observations, therefore if the continuum varies between WHT observations, the BLR ought to respond to this change. So in the event of the central engine becoming more luminous, the BLR would originate further from the source, and correspondingly the line velocity width would be smaller. Since for the intrinsic case we analyse the spectra as observed, and for the extrsinsic case we scale to the quiescent state (which alters the continuum luminosity, but not the linewidth), it is possible that the \M_BH estimates would be more closely grouped between epochs for the favoured model. However, we see no evidence for such an effect -- the scatter arising from the method is too great.

The absorption features in J1422$+$0140 are consistent with an intervening system. Whilst the doublet on the blue wing of the \Mgii profile (at 2750 \AA) is likely \Mgii absorption in an outflowing component intrinsic to the AGN, the second doublet seen at 2500 \AA \ in Fig.\ \ref{fig:masses} is either \Mgii absorption in an extreme outflow (at $0.1 c$) or more probably in an intervening system at $z \simeq 0.855$. This could be a signature of the lens host galaxy, in scenario (B). More discussion and interpretation of these features is presented in \cite{bruce16}.

\section{X-ray Spectrum and Variability} \label{sec:xray}

\begin{table*}
\caption{\small The X-ray spectra model properties. Photoelectric absorption components for both the Milky Way (fixed) and the host galaxy (free) are modelled. In some observations, the host galaxy absorption is poorly constrained, due to both the redshift and number of counts. The EPIC count errors are 1 $\sigma$, and the errors on the model parameters are the 90 per cent confidence limits, in line with convention in X-ray astronomy. We also quote $\Gamma$ and its uncertainty to two decimal places. The exposure time is the full observation time on target, including time that was determined by the \xmm pipeline to be `bad'.}
\small
\centering
\begin{tabular}{ccccccccc}
\hline
Name       & Obs. UT           & Exp. Time   & \xmm EPIC Cts       & $N_{\rm H, \, MW}$ 
           & $N_{\rm H,\,int}$ & $\Gamma$              & Norm (1 keV)          & \rchi    \\
           &                   & (s)         &                     & ($\times 10^{20}\,$cm$^{-2}$)
           & ($\times 10^{20}\,$cm$^{-2}$) & & ($\times 10^{-6}$ Ph.\ $/$ cm$^{2}$ s keV) &    \\
\hline
J0312$+$1836  & 2013-08-12        & 35 100      & 1130$\pm$40         & 8.02   
              & 0$^{+6}_{-0}$     & 2.26$^{+0.14}_{-0.12}$& 24.6$^{+1.9}_{-1.5}$  & 0.99  \\
              & 2014-01-22        & 29 500      & 205$\pm$16          & 8.02 
              & 10$^{+40}_{-10}$  & 2.46$^{+0.57}_{-0.36}$& 16$^{+7}_{-4}$        & 1.70  \\
\vspace{-0.2 cm} \\
J1422$+$0140  & 2014-02-06        & 34 500      & 1320$\pm$40         & 2.57 
              & 0$^{+14}_{-0}$    & 1.73$^{+0.18}_{-0.09}$& 15.6$^{+2.2}_{-0.9}$  & 1.12  \\
\vspace{-0.2 cm} \\
J1519$+$0011  & 2014-01-27        & 23 700      & 4250$\pm$70         & 4.57 
              & 0$^{+2}_{-0}$     & 1.96$^{+0.06}_{-0.05}$& 90$^{+4}_{-3}$        & 1.01 \\
              & 2014-02-10        & 28 000      & 4850$\pm$70         & 4.57   
              & 0.0$^{+0.9}_{-0.0}$& 2.04$\pm0.05$        & 80$^{+2}_{-2}$        & 1.05 \\
\vspace{-0.2 cm} \\
J2232$-$0806  & 2013-12-14        & 29 500      & 9300$\pm$100        & 4.11  
              & 0.0$^{+0.4}_{-0.0}$& 2.21$\pm0.04$        & 284$\pm$7             & 1.10  \\
\hline

\end{tabular}
\label{tab:xray}
\end{table*}

We first carry out an analysis of the X-ray spectrum only, by fitting an absorbed power-law to the X-ray spectral data. This model is simple, but at the redshift of these objects we anticipate that it will be appropriate, given the relatively low count numbers. We incorporate absorption components attributable to both the Milky Way (fixed) and the host galaxy (free). We are able to test for any statistically significant variability between observations in J0312$+$1836 and J1519$+$0011 -- the two objects for which we have useable data from two epochs. Milky Way \nh \ values come from the Leiden/Argentine/Bonn survey \citep{kalberla05}. We use the \xspec spectral analysis package, and a Levenberg-Marquardt minimisation routine for all fitting.

The X-ray exposure times and count values are shown in Table \ref{tab:xray}, together with the fitted parameters and 90 per cent confidence limits.

The intrinsic \nh \ in each object is low. This may be because our HVAs are at moderate redshifts, so only the tail of the photoelectric absorption profile is sampled by the X-ray spectrum, and have relatively low count numbers, increasing the uncertainty on $N_{\rm H,\,int}$. A specific example of this is evident in the second observation of J0312$+$1836, where the observation with just $\sim 200$ counts shows a broad 90 per cent confidence limit on $N_{\rm H,\,int}$.

We observe a range in power-law slopes ($\Gamma$). Flat power law slopes are characterised by $\Gamma = 2$, as seen in J1519$+$0011, soft slopes by $\Gamma > 2$, as seen in J0312$+$1836 ($\sim 3 \sigma$ significance) and J2232$-$0806 ($\sim 9 \sigma$ significance), and hard slopes by $\Gamma < 2$, which we observe in J1422$+$0140 to a smaller ($\sim 2 \sigma$) significance.

There is evidence for statistically significant ($\simeq \! 4.5 \sigma$) variability between the observations of J1519$+$0011, despite these observations being just 14 days apart. We see that whilst $\Gamma$ is consistent between the two observations, the normalisation has decreased, indicating that the object faded between observations. The optical light curve over the same period (Fig.\ \ref{fig:lightcurves_xmm}) does not show a significant change, possibly due to larger scatter and insufficient temporal sampling. We could therefore be seeing a differentially fading X-ray component in J1519$+$0011, or this could be due to fast X-ray variability commonly seen in AGN (\eg \citealt{gierlinski08}, \citealt{parker15}). It is possible the significance of this event could be exaggerated by residual background activity not fully taken into account in the reduction process.

We also test for short-term variability during the observations of each target. Given the relatively low count numbers, this is done by fitting the short term light curve ($\sim 100$ count bins) with a constant. We only use the `good' on-target time, where background activity was low, and use the full 0.2$-$12 keV range. Deviation from unity in the \rchi \ fitting statistic can provide evidence for such variability, but we do not observe this in any object. There is therefore no suggestion of statistically significant, short-term X-ray variability in any of the objects.

\begin{table*}
\caption{\small The optimum fitted parameters for the various SED models. Uncertainties quoted are the 90 per cent confidence limits, as is conventional in X-ray astronomy, and are estimated using the Fisher matrix. As such, they are only indicative of the true measurement error. Columns are as follows: (1) object name, (2) the lens factor, if applied -- this is a constant factor that multiplies the model flux at all energies, and derives from $\Delta m$ in Table \ref{tab:mags}, (3) log bolometric luminosity [$\log$ (erg s$^{-1}$)], (4) log luminosity density at 2500 \AA \ [$\log$ (erg s$^{-1}$ Hz$^{-1}$)], (5) log luminosity density at 2 keV [$\log$ (erg s$^{-1}$ Hz$^{-1}$)], (6) $\alpha_{\rm OX}$ spectral index (\eg \citealt{lusso10}), (7) intrinsic ($B-V$) extinction [mag], (8) reduced mass accretion rate [$\dot{M}_{\rm Edd}$], (9) coronal radius [$R_{\rm g}$], (10) outer disc radius [$R_{\rm g}$], (11) PLT spectral index, (12) \rchi \ fitting statistic.}
\small
\centering
\begin{tabular}{cccccccccccc}
\hline
Name & $f_{\rm lens}$    & $L_{\rm bol}$    & $L_{2500\rm \mathring{A}}$& $L_{2 \rm keV}$  
   & $\alpha_{\rm OX}$   & E$(B\!-\!V)$     & $\dot{m}$        & $r_{\rm cor}$ 
   & $r_{\rm out}$       & $\Gamma$         & \rchi \\
(1)& (2)                 & (3)              & (4)              & (5)
   & (6)                 & (7)              & (8)              & (9)
   & (10)                & (11)             & (12)              \\
\hline

\multicolumn{9}{l}{\textbf{Model A: As observed.} \M_BH $=M_{\rm BH,observed}$ in Table \ref{tab:masses}.}\\

\vspace{-0.3 cm} \\

J0312$+$1836 & N/A & 45.87$\pm$0.12 & 30.25 & 26.44 & 1.46 & 0.07$\pm$0.02 & 0.12$\pm$0.03 & 13$\pm$2 & 93$\pm$8 & 2.24$\pm$0.11 & 1.83 \\

J1422$+$0140 & N/A & 45.72$\pm$0.10 & 30.04 & 26.53 & 1.35 & 0.06$\pm$0.03 & 0.12$\pm$0.03 & 20$\pm$13 & $>$1000 & 1.75$\pm$0.10 & 1.15 \\

J1519$+$0011 & N/A & 45.8$\pm$0.3   & 30.10 & 26.50 & 1.38 & 0.16$\pm$0.06 & 0.09$\pm$0.06 & 13$\pm$4 & 49$\pm$3 & 2.00$\pm$0.03 & 1.11 \\

J2232$-$0806 & N/A & 45.30$\pm$0.17 & 29.49 & 26.31 & 1.22 & 0.00$\pm$0.04 & 0.10$\pm$0.04 & 30$\pm$20 & 120$\pm$6 & 2.20$\pm$0.05 & 1.11 \\

\vspace{-0.2 cm} \\

\multicolumn{9}{l}{\textbf{Model B: Descaled.} \M_BH $=M_{\rm BH,faint}$ in Table \ref{tab:masses}.}\\

\vspace{-0.3 cm} \\

J0312$+$1836 & 4.53 & 45.27$\pm$0.14 & 29.57 & 25.78 & 1.46 & 0.07$\pm$0.03 & 0.08$\pm$0.03 & 12$\pm$2 & 134$\pm$13 & 2.24$\pm$0.11 & 1.83 \\

J1422$+$0140 & 6.43 & 45.0$\pm$0.2 & 29.25 & 25.72 & 1.35 & 0.06$\pm$0.05 & 0.08$\pm$0.04 & 17$\pm$6 & $>$1000  & 1.75$\pm$0.09 & 1.15 \\

J1519$+$0011 & 4.66 & 45.1$\pm$0.3 & 29.40 & 25.83 & 1.37 & 0.15$\pm$0.07 & 0.05$\pm$0.04 & 13$\pm$4 & 71$\pm$5 & 2.00$\pm$0.03 & 1.11 \\

J2232$-$0806 & 2.25 & 44.95$\pm$0.09 & 29.14 & 25.96 & 1.22 & 0.000$\pm$0.019 & 0.065$\pm$0.014 & 29$\pm$14 & 119$\pm$7 & 2.20$\pm$0.05 & 1.11 \\

\hline

\end{tabular}
\label{tab:seds}
\end{table*}

\begin{figure*}
\centering
\includegraphics[trim=0cm 1cm 0cm 2.5cm, clip=true, width = \textwidth]{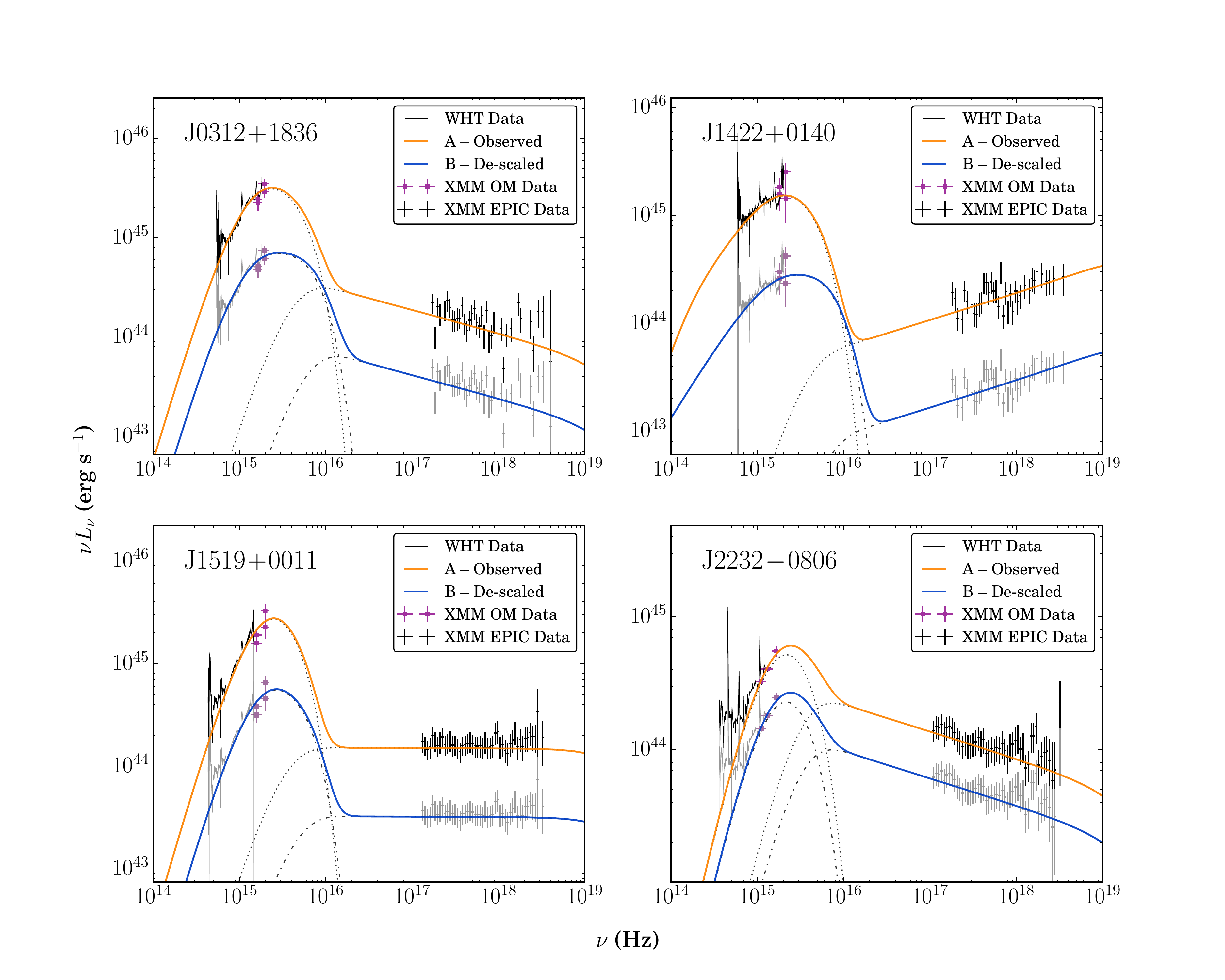}
\caption{\small Modelled SEDs for both scenarios. In the extrinsic variability scenario, we assume that the observed data is not representative of the normal state, and that the representative flux is lower than observed.
In these plots, we show the intrinsic SED in each case; for model B, we therefore scale the data down to the implied intrinsic flux, shown in grey. Model constituent components (AD and PLT) are shown by the dotted and dash-dotted lines.}
\label{fig:seds}
\end{figure*}

\section{Broad band spectral energy distribution}  \label{sec:seds}

\subsection{SED construction} \label{subsec:sed_cons}

In the optical regime, we use the WHT spectrum that was observed closest to the \xmm observation date, and normalised in the $r$ band to the \xmm observation epoch (see Section \ref{sec:lightcurves}). From this spectrum we define bins expected to be free from emission features (\Feii , emission lines and the Balmer continuum) as being representative of the continuum flux level. This is discussed in greater detail in Collinson et al.\ (2015, 2017). \nocite{collinson15} \nocite{collinson17}

The \xmm OM makes UV photometric measurements in a range of optical and UV bands (see \citealt{mason01}). The OM data is expected to be biased high by the presence of emission features in the bandpass of each filter. We correct for this by estimating the flux surplus using the \cite{vandenberk01} quasar template and derived power-law continuum, combined with the effective bandpass of each OM filter at each redshift. By integrating the template over the OM bandpasses to simulate the total flux measured (including emission features) and then integrating the power-law continuum over the same bandpasses to simulate the true continuum level, we estimate the factor by which emission features increase the observed flux in each band. We then scale the OM photometry to correct for this difference, so the OM data represents the continuum flux level. \cite{elvis12} demonstrated that variation in the equivalent width of the Lyman-$\alpha$ emission feature made utilising a single value from a template unreliable. Hence we did not include the UVM2 band in J0312$+$1836 in our modelling, as it lay directly on top of Ly-$\alpha$. The correction factor, $f_{\rm corr,OM}$, was typically $0.8 \lesssim f_{\rm corr,OM} \lesssim 0.9$, dependent on the filter and redshift of each object.

We do not use data from the Two-Micron All-Sky Survey (2MASS), UKIRT Infrared Deep Sky Survey (UKIDSS) or the \textit{Wide-field Infrared Survey Explorer} (\wise) to extend the SED into the IR, or the \textit{Galaxy Evolution Explorer} (\galex) to extend into the UV, as we require quasi-simultaneous data. Also, our objects were often below the detection threshold of these surveys, at the time of observation.

\subsection{SED modelling} \label{subsec:sed_mod}

In this section we model the broad-band SED of each object. We apply the AGN SED model of \cite{done12} -- \optxagnf \ -- to each object, fitting to data from \xmmn (including the onboard optical monitor -- OM) and continuum regions of the WHT spectrum. This model comprises three components; accretion disc (AD), soft X-ray excess (SX) and power-law tail (PLT), see \cite{done12} for a more detailed explanation. Importantly the model conserves energy between these components, deriving energy from the release of gravitational potential energy by the accreting gas. We also use multiplicative components to model the extinction and soft X-ray absorption of the MW.

The attenuation variables are the Milky Way extinction, E($B\!-\!V$)$_{\rm MW}$, (fixed using values from \citealt{schlafly11}), the intrinsic reddening, E($B\!-\!V$)$_{\rm int}$ (free), and, as in Section \ref{sec:xray}, the hydrogen column densities for the photoelectric absorption -- $N_{\rm H, MW}$ and $N_{\rm H, int}$. Following the procedure described in \cite{collinson15}, we test SED models both with and without attenuation attributable to the AGN host galaxy. However, as found in that work (see also \citealt{capellupo15} and \citealt{castello-mor16}) the model with host galaxy reddening/absorption produces a better model fit in all cases, when allowing for the additional free parameters. We will therefore consider SED models that include these components from this point onwards. We tested two extinction curve models for the intrinsic reddening: Milky Way (MW) and Small Magellanic Cloud (SMC). All objects were better fit with the MW extinction curve, but the difference in \rchi \ was marginal between the two cases in J2232$-$0806.

Due to the redshift of our objects, and the quality of our X-ray data, we do not well-sample the SX component with the \xmm data. In \cite{collinson15}, it was found that the inclusion of the SX was justified as it was partially sampled in high-mass (\M_BH$>10^9$) objects. However, that sample was of higher redshift ($z>1.5$), and consequently higher \M_BH \citep{mclure04}. For the current sample, we initially applied two versions of the model; one with a dominant SX (initially contributing 70 per cent of the reprocessed energy -- \eg \citealt{done13}, \citealt{collinson15} -- which was then allowed to vary), and one with no SX. We found that in some cases the model including the SX was not physical (\eg because all AD energy was reprocessed by the SX and PLT). J2232$-$0806 did show evidence for an improved fit with the SX component (\rchi \ of 1.07 compared to 1.11 without the SX component), however, the difference in end result was small. Therefore, in the interest of simplicity and consistency we opted to use the simplified, two component (AD and PLT) version of \optxagnf.

Using an energetically self-consistent accretion model allows us to compare the two hypotheses (high-state normal or low-state normal, see Section \ref{subsec:mechs_thispaper}) for the increase in brightness of these HVAs. If the change is intrinsic, then the spectral data will represent the energy flux of a conventional AGN that has seen an increase in accretion rate. In the event that the increment in brightness is due to an extrinsic effect, then the intrinsic flux of the AGN must be a significant factor smaller than that we observe, corresponding to a smaller mass accretion rate.

We test these two scenarios, by producing corresponding models for each object:

\begin{enumerate}
\item[{\textbf A.}] As observed. The model will be fitted to the data with no additional factors applied.
\item[{\textbf B.}] Change is extrinsic. We de-scale the SED at all wavelengths to match the SDSS low-state.
\end{enumerate}

This enables us to contrast the intrinsic and extrinsic scenarios and look for evidence of unusual effects, by comparing the inferred properties with those of larger AGN samples. As discussed in Section \ref{sec:bhmasses}, we have estimated \M_BH for each of these situations, and we will therefore use a different \M_BH estimate for the two scenarios, which is fixed in the fitting. The resulting model properties are tabulated in Table \ref{tab:seds} and the SEDs for both cases are shown in Fig.\ \ref{fig:seds}. The only fitted property we do not tabulate is $N_{\rm H, int}$, as these are all at or close to zero, as in Table \ref{tab:xray}.

\section{Discussion}  \label{sec:discussion}

\subsection{SED Model} \label{subsec:discussion:sed}

\begin{figure}
\centering
\includegraphics[trim={0cm 0.5cm 0cm 0.5cm},clip, width = 0.5 \textwidth]{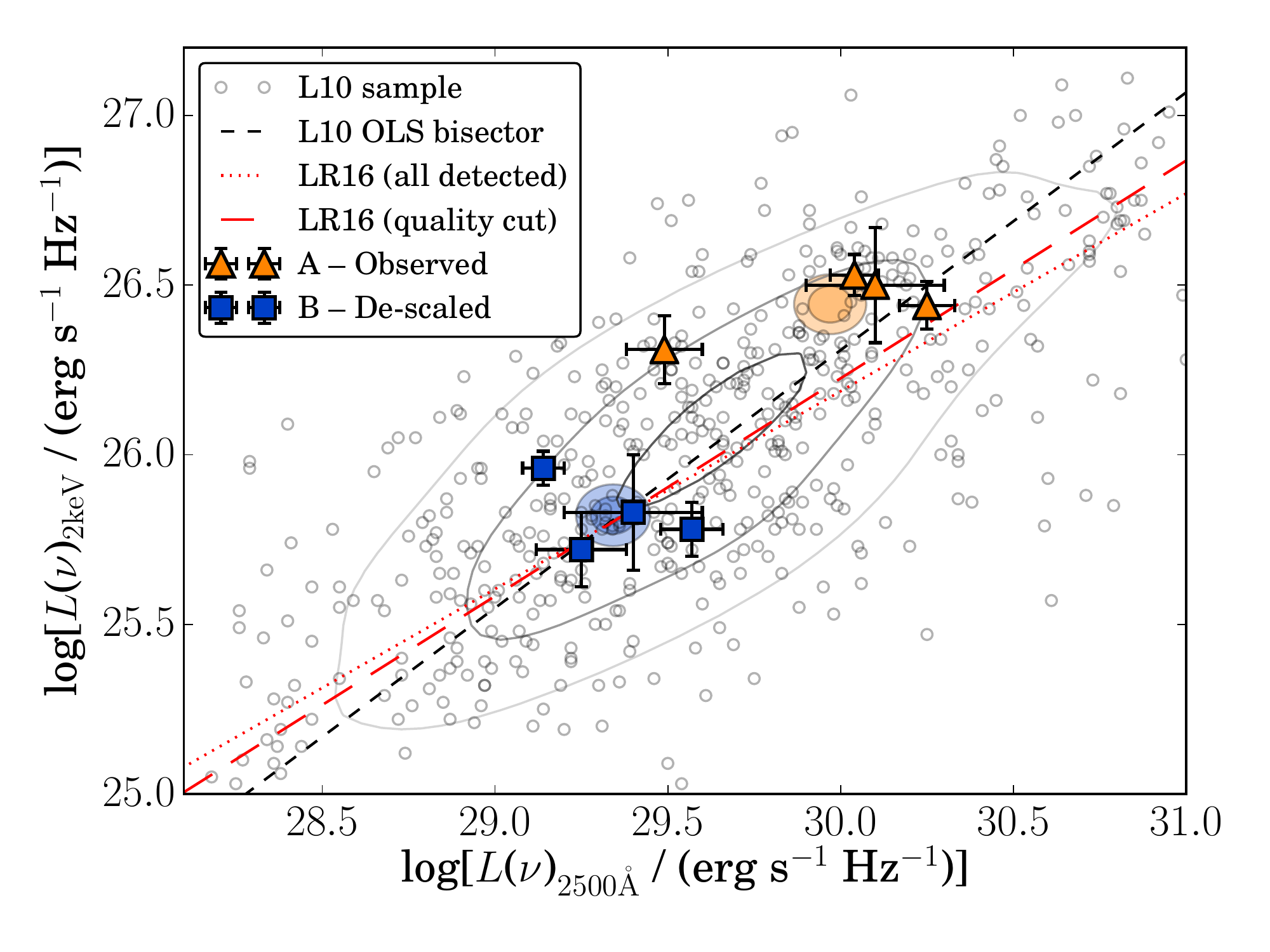}
\caption{\small The L10 sample, $L(\nu)_{2 \rm keV}$ against $L(\nu)_{2500 \rm \mathring{A}}$ with our sample (blue and orange) overplotted. LR16 relations are also shown. The contours illustrate the distribution of the L10 sample, and the ellipses show the sample centroids ($1$ and $2 \sigma$ significance levels).}
\label{fig:lusso6}
\end{figure}

\begin{figure}
\centering
\includegraphics[trim={0cm 0.5cm 0cm 0.5cm},clip, width = 0.5 \textwidth]{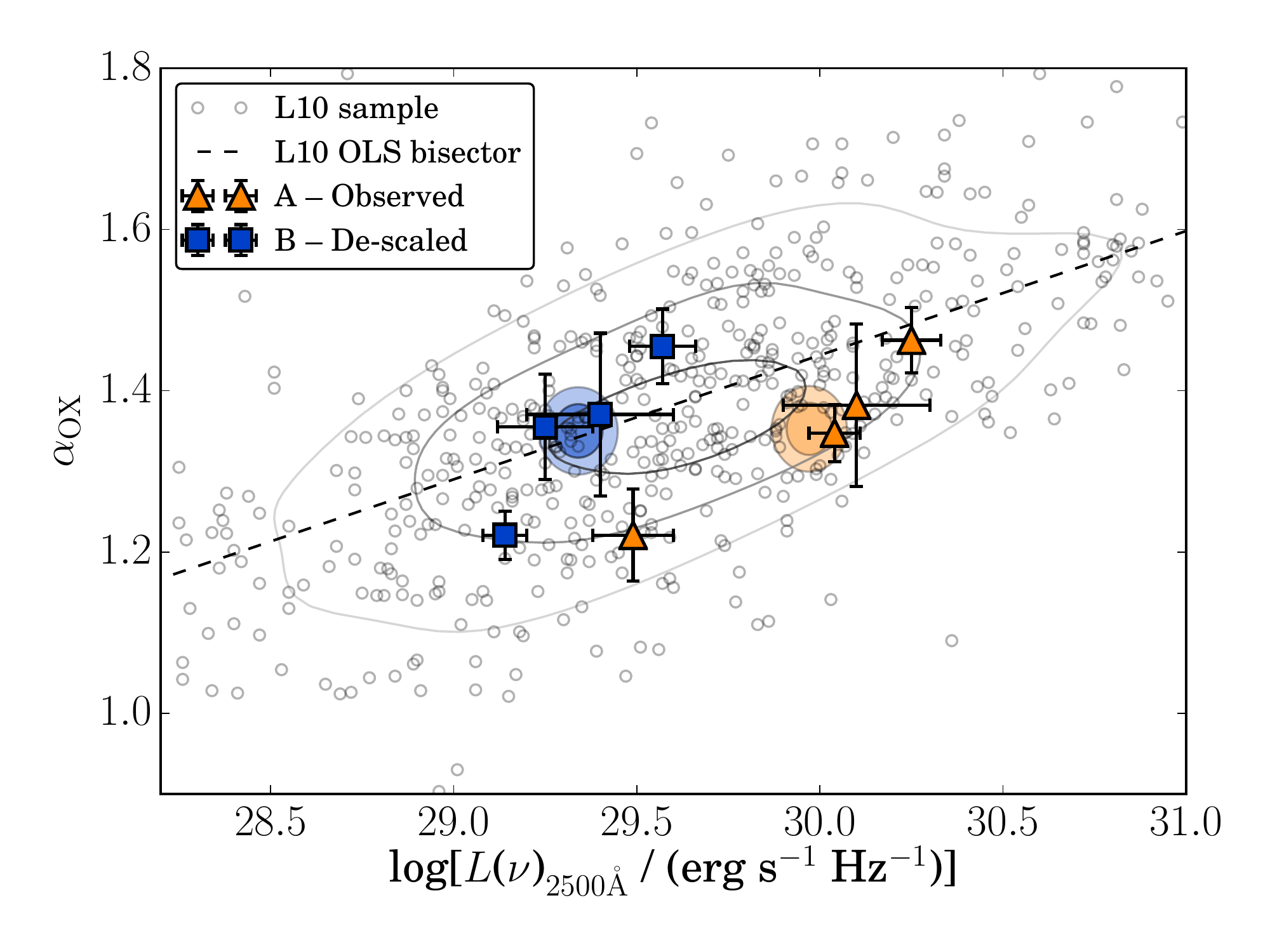}
\caption{\small The L10 sample, $\alpha_{\rm OX}$ against $L(\nu)_{2500 \rm \mathring{A}}$ with our sample (blue and orange) overplotted.}
\label{fig:lusso7}
\end{figure}

\begin{figure}
\centering
\includegraphics[trim={0cm 0.5cm 0cm 0.5cm},clip, width = 0.5 \textwidth]{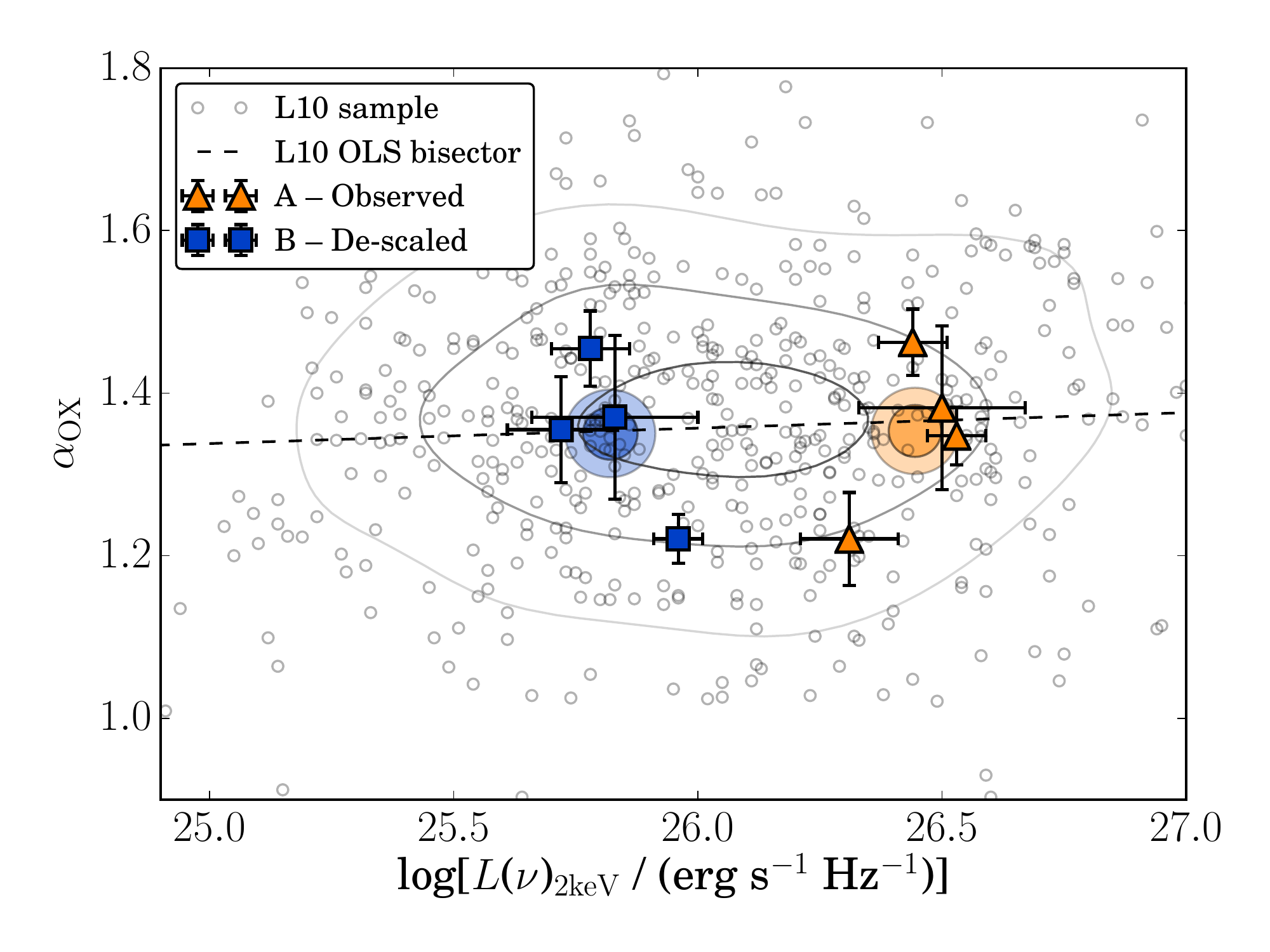}
\caption{\small The L10 sample, $\alpha_{\rm OX}$ against $L(\nu)_{2 \rm keV}$ with our sample (blue and orange) overplotted.}
\label{fig:lusso8}
\end{figure}

Figs.\ \ref{fig:lusso6}, \ref{fig:lusso7} and \ref{fig:lusso8} show the the results from our model fitting, in the context of a large sample of normal AGN (\citealt{lusso10}, hereafter L10).

L10 presented SEDs for a large, X-ray selected sample of 545 Type 1 AGN, drawn from the \xmmn Cosmic Evolution Survey (COSMOS) sample. They estimated the multi-waveband SEDs for their sample using multiple polynomial interpolations and extrapolated power-laws  through their data in log($\nu L_{\nu}$) space. This method does not apply any physical considerations to the procedure, but adheres to known constraints  (\eg range of the PLT). We compare our calculated values for $L_{2500 \rm \mathring{A}}$, $L_{2 \rm keV}$ and $\alpha_{\rm OX}$ with those in the L10 sample. $\alpha_{\rm OX}$ is an often-used measure of the relative X-ray loudness of an AGN, and is defined as:
\begin{equation}
\alpha_{\rm OX} = -\frac{\log(L_{2 \rm keV}/L_{2500 \rm \mathring{A}})}{2.605} .
\end{equation}

We plot the linear best fit relations derived in L10 using the method of \cite{isobe90}. The ellipses show the $1 \sigma$ and $2 \sigma$ error regions for the HVA sample centroid of each tested model. These regions are calculated using a Monte Carlo method, similar to that described in Section \ref{sec:lightcurves}. The central 68 and 95 per cent of these centroid distributions are an indication of the $1 \sigma$ and $2 \sigma$ error boundaries respectively. To guide the eye, we also overlay contours from a bivariate Gaussian kernel-density estimate of the L10 sample distribution.

In Fig.\ \ref{fig:lusso6} we also show the linear relations derived in \cite{lusso16}, hereafter LR16. The LR16 sample comprises 2153 AGN detected in both SDSS and the \xmm serendipitous source catalog. They calculate $L_{2 \rm keV}$ from the \xmm EPIC total energy fluxes, assuming a constant photon index and neutral hydrogen column. $L_{2500 \rm \mathring{A}}$ values were provided in the \cite{shen11} catalog. This approach differs from that in L10, but there is agreement between the studies with respect to the $L_{2500 \rm \mathring{A}}$--$L_{2 \rm keV}$ relation. LR16 show that by applying various quality cuts, the dispersion of this relation drops significantly. In Fig.\ \ref{fig:lusso6}, the relation derived from the full LR16 sample is shown by the red dotted line, and the relation emerging from the best-quality subsample (743 objects) by the red long-dashed line. 

The most direct result of this analysis can be seen in Fig.\ \ref{fig:lusso7}. For their observed optical--UV luminosities, the four HVAs are systematically X-ray loud -- they all lie under the trend line of Fig.\ \ref{fig:lusso7}. Given the trend of \a_OX with luminosity, this could be because the luminosities are systematically over-estimated. If we apply the de-scaling factor estimated from our scenario B model fits, the HVAs fall on the trend line. However, although the result is formally significant at $>2 \sigma$, given that we have only four data points, it is hard to be confident of this result. A larger sample of HVAs is clearly desirable.

The comparison with L10 and LR16 ought to be reasonable; whilst our approaches for calculating the SEDs differ, we are primarily concerned with $L(\nu)_{2500 \rm \mathring{A}}$ and $L(\nu)_{2 \rm keV}$ which are well-sampled, and thus well-defined by our various approaches. Uncertainty may arise in $L(\nu)_{2500 \rm \mathring{A}}$ due to differing means of correcting for intrinsic reddening. Owing to the limited quality of their data, L10 make only simple corrections for host galaxy reddening. Unfortunately, AGN samples that utilise more advanced SED models that address these limitations either cover limited redshift ranges (\eg \citealt{jin12_1}), or lack X-ray data (\eg \citealt{capellupo15}), in addition to being much smaller. However, LR16 explore the effects of a number of quality criteria (including reddening) in their large sample of AGN. They find that whilst the dispersion of the $L(\nu)_{2500 \rm \mathring{A}}$--$L(\nu)_{2 \rm keV}$ relation is dependent on such criteria, the relation itself is not. Therefore our findings should not be affected by this source of uncertainty.

We may also consider the average AGN templates of \cite{jin12_1} as being representative of the archetypal AGN SEDs at different accretion rates, as they use the same model as in this paper, applied to a larger, more local AGN sample. In terms of \a_OX, as a sample our HVAs appear to most closely resemble the average SED for objects with moderate linewidth. However, even within four objects, we see a broad range of X-ray spectral shapes ($1.8 \lesssim \Gamma \lesssim 2.2$) that almost covers the full range of SEDs observed by \cite{jin12_1}. For instance J0312$+$1836 shows an unusually soft X-ray spectrum, compared to the average SED in \cite{jin12_1}, for an object at the accretion rates predicted in either case. To this end, a more conclusive result would require a larger sample of HVAs with X-ray data.

In both situations we generally predict moderate intrinsic reddening (E$(B-V)\simeq0.07$, $n=4$ objects). This is a little higher than typical extinction values determined in \cite{collinson15} (mean E$(B-V)\simeq0.04$, $n=11$ objects) and \cite{capellupo15} (mean E$(B-V)\simeq0.02$, $n=30$ objects). Our sample only comprises four objects, so it is impossible to say this is a true trend among HVAs. But higher-than-usual extinctions may be expected if there are two galaxies (\ie both AGN host galaxy and lens host galaxy) extinguishing the optical/UV light. These reddening curves would be at different redshifts (\ie the host and lens host galaxy), but we would need better data coverage and signal-to-noise to deconvolve these two contributions, especially as the total $(B-V)$ extinction we measure is $< 0.1$ mag.

The intrinsic extinction parameter is dependent on the slope of the optical spectrum. Other factors that could affect the spectral slope are the outer AD radius, and host galaxy contamination (potentially with an additional lens host galaxy, if these are indeed microlensed AGN). We do not model the host galaxy component in this study, as we do not expect the host galaxy to make a significant contribution to the SED blueward of \Hb (\citealt{shen11}, \citealt{collinson17}). We tested a version of the model where the outer disc radius was fixed at 1000 $R_{\rm g}$, but in 3 objects the fit was marginal.

The evidence we see for a fading X-ray component (Section \ref{sec:xray}) in J1519$+$0011 is interesting, as if it continued to fade at that rate (\ie significantly faster than the optical), it could suggest differential magnification of the source, and provide a probe of the corona size. To confirm this would require follow up observations with \xmmn .

\subsection{Additional Uncertainties} \label{subsec:discussion:add}

There is good agreement between the scaled optical (WHT) data and the corrected (\xmm OM) UV data, suggesting that it is reasonable both to normalise the WHT spectrum using the optical light curve, and to correct the OM data for emission feature contamination using the \cite{vandenberk01} template and power-law continuum.

It should be noted that in order to estimate \M_BH in the extrinsic variability case (Model B), we scaled each source to the brightness level of the quiescent state, which we assumed to be represented by the SDSS magnitude. However, in cases where the SDSS observations are dominated by the host (or foreground) galaxy, the AGN flux could be fainter. In this case, \M_BH would be smaller, meaning an AD that peaks further into the UV, with higher mass accretion rates. Our \M_BH for the extrinsic case should therefore be considered upper limits.

A final caveat is that if the BHs are spinning, the radius of last stable circular orbit is reduced and a co-rotating AD can extend closer to the event horizon, shifting the AD peak bluewards. Unfortunately our data are insufficient to make a judgement of the BH spin (see \citealt{collinson17}).

\section{Summary and conclusions}  \label{sec:conclusions}

We have presented results from X-ray observations of four extremely variable AGN (termed `hypervariable' AGN, or HVAs) discovered in the Pan-STARRS database (Lawrence et al.\ 2012, 2016). To explain this variability we consider two distinct scenarios -- (A) and (B). In (A), this change is intrinsic (\eg caused by a large increase in mass accretion rate), and in (B), the flux has been increased by some extrinsic factor, such as for example foreground microlensing. We explore these two scenarios through an analysis of their SEDs.

We have estimated the magnitudes of each AGN in both faint and bright states, and also at the epoch of the \xmmn observation, using optical photometry light curves. We then use optical spectra from WHT/ISIS to estimate \M_BH from the profiles of the broad emission lines \Mgii and \Ha. For (A) we analyse the spectra as observed, and for (B), we scale them to the faint (assumed to represent the intrinsic) state.

We then fit an energy-conserving, broad band SED model to the multi-wavelength data for each object. This approach allows us not only to characterise the energetics for each scenario, but also constrains the accretion flow properties, including the SED shape, which is dependent on the mass accretion rate.

We compare the properties of the models for each of the four objects, in both scenarios (A and B) with the \cite{lusso10} (L10) and \cite{lusso16} samples. Our four HVAs show distinct groupings of (A) and (B) in SED shape versus luminosity parameter space. In scenario (A), we see evidence that the AGN are X-ray loud for their apparent UV luminosities, whereas in (B), their X-ray loudness seems normal.

This provides an important additional diagnostic of the expected arrangement of these HVAs. With a larger sample of X-ray observed HVAs, we hope to be able to increase the significance of our findings.

\section*{Acknowledgements}  \label{sec:acknowl}
The authors would like to thank the referee for constructive comments. We appreciate the help of Marianne Vestergaard, who provided the UV \Feii templates from \cite{vestergaard01}, and JSC is forever grateful to Edward Shaw for everything he contributed. JSC is funded by Science and Technology Facilities Council (STFC) grant ST/K501979/1 and AB by the Edinburgh PCDS scholarship. MJW and CD are supported by STFC grant ST/L00075X/1.

This research is based on observations obtained with \xmmn, an ESA science mission with instruments and contributions directly funded by ESA Member States and NASA.

The Pan-STARRS 1 Survey has been made possible through contributions of the Institute for Astronomy; the University of Hawaii; the Pan-STARRS Project Office; the Max-Planck Society and its participating institutes, the Max Planck Institute for Astronomy, Heidelberg and the Max Planck Institute for Extraterrestrial Physics, Garching; The Johns Hopkins University; Durham University; the University of Edinburgh; Queen’s University Belfast; the Harvard-Smithsonian Center for Astrophysics; and the Las Cumbres Observatory Global Telescope Network, Incorporated; the National Central University of Taiwan; and the National Aeronautics and Space Administration under Grant No.\ NNX08AR22G issued through the Planetary Science Division of the NASA Science Mission Directorate.

The William Herschel Telescope is operated on the island of La Palma by the Isaac Newton Group. The Liverpool Telescope, also on the island of La Palma, is operated by Liverpool John Moores University with financial support from the UK Science and Technology Facilities Council. Both are in the Spanish Observatorio del Roque de los Muchachos of the Instituto de Astrof\'{i}sica de Canarias.

Funding for SDSS-III has been provided by the Alfred P. Sloan Foundation, the Participating Institutions, the National Science Foundation, and the U.S. Department of Energy Office of Science. The SDSS-III web site is http://www.sdss3.org/.

Finally, we are grateful to the contributors to the {\sc python} programming language, and HEASARC, for software and services.

\small{
\bibliography{Collinson}
\bibliographystyle{mn2e}
}

\label{lastpage}

\end{document}